\documentclass{article}

\usepackage{fullpage}
\usepackage{graphicx}
\usepackage{amsmath}
\usepackage{amsthm}

\usepackage[ruled]{algorithm2e}
\usepackage{url}
\usepackage{tikz}
\usepackage{pgf}
\usepackage{caption}
\usepackage{subcaption}
\usetikzlibrary{backgrounds}
\usetikzlibrary{arrows,automata}
\usetikzlibrary{positioning}
\usetikzlibrary{fit}
\usetikzlibrary{decorations.pathmorphing} 
\usetikzlibrary{shapes,shadows,calc}
\providecommand{\SetAlgoLined}{\SetLine}
\usepackage{mathtools}
\usepackage{multirow}
\DeclareMathOperator*{\argmax}{arg\,max}

\usepackage[round]{natbib}

\begin{document}

\title{DPM: A State Space Model for Large-Scale Direct Marketing}
\author{ Yubin Park \\ Accordino Health, Inc. \and Rajiv Khanna \\ UT Austin \and Joydeep Ghosh \\ UT Austin   \and Daniel Mihalko \\ USAA  }

\maketitle

\begin{abstract}
We propose a novel statistical model to answer three  challenges in direct marketing: 
\textit{which channel} to use, \textit{which offer} to make, and \textit{when} to offer.
There are several potential applications for the proposed model; for example, developing personalized marketing strategies and monitoring members' needs.
Furthermore,  the results from the model can complement and can be integrated with other existing models.

The proposed model, named Dynamic Propensity Model, is a latent variable time series model that utilizes both marketing and purchase histories of a customer. 
The latent variable in the model represents the customer's propensity to buy a product. The propensity derives from purchases and other observable responses.
Marketing touches increase a member's propensity,  
and propensity score attenuates and propagates over time as governed by data-driven parameters.
To estimate the parameters of the model,  a new statistical methodology has been developed.
This methodology makes use of particle methods with a stochastic gradient descent approach, 
resulting in fast estimation of the model coefficients even from big datasets.
The model is validated using six months' marketing records from one of the largest insurance companies in the U.S. 
Experimental results indicate that the effects of marketing touches vary depending on both channels and products.
We compare the predictive performance of the proposed model with lagged variable logistic regression.
Limitations and extensions of the proposed algorithm are also discussed.
\end{abstract}

\section{Introduction}
Direct marketing is a form of data-driven advertising that markets straight to potential consumers through various marketing channels.
\cite{DMA2010} reported that in 2010, business and non-profit organizations spent $\$153$ billion on direct marketing, which is approximately $54\%$ of all advertisement expenditures in the United States.
The scope of direct marketing channels has expanded from traditional direct mails to targeted online display ads, search, and social media sites, and the sizes of consumer databases have significantly increased. 
In recent years, web-service companies, such as Google and Yahoo, have applied large-scale data mining and machine learning techniques to online ad optimization e.g. \cite{perlich2013} detail one such system,
\cite{Yan2009} empirically showed that behavioral targeting improves click-through rates,
\cite{Gupta2012} applied matrix factorization for display advertisement matching, \cite{khanna2013} developed specialized methods for large scale modeling for targetted ads on hadoop,
and in a recent KDD Workshop on Data Mining for Advertising, \cite{Shen2008} covered technological advances in online direct marketing. 
The landscape of direct marketing is dynamically changing and expanding to new applications. 

Two statistical models are particularly popular for direct marketing: uplift modeling and RFM (Recency, Frequency, and Monetary value analysis).
Uplift modeling, also known as incremental modeling or net-lift modeling, aims to measure the effectiveness of advertisement through randomized control and test groups \citep{Radcliffe1999,Lo2002}.
Although uplift modeling can provide strong return on investment cases for some applications, a large number of control and test samples are needed to estimate extremely low response rates e.g. insurance purchases.
While uplift modeling predicts the difference that a marketing touch makes, RFM is a marketing-agnostic summary statistics for describing customers \citep{Fader2005}.
In the RFM modeling, a customer is represented using three variables; the time of the most recent purchase, the frequency of purchases, and average spending per purchase. 
\cite{McCarty2007} compared the predictive performances of RFM, CHAID, and logistic regression, and concluded that RFM was comparable to the other two methods.
However, since RFM is marketing-agnostic, high-lift customers, whose purchase behaviors are significantly affected by marketing touches, tend to be neglected when RFM is used as the primary targeting method.

State space models, although not popular in direct marketing literature, can potentially address three main challenges in direct marketing: \textit{which channel} to use, \textit{which offer} to make, and \textit{when} to offer.
A state space model, also known as dynamic linear model, is a latent variable time series model that represents a physical system as a set of input, output, and (latent) state  variables.   
In direct marketing, the input and output refer to the marketing touch and the response respectively, and the state can be interpreted as propensity to purchase a product. 
Although state space models have been widely adopted in diverse academic domains such as machine learning, economics, and finance (see Section~\ref{sec:ssmest}), there have been few applications in direct marketing to the best of our knowledge.  
\cite{Dekimpe2000} provide two reasons for the poor adoption of time series modeling in marketing: lack of adequate data sources and lack of easy-to-use times series softwares for marketers.
\cite{Pauwels2004} summarized potential challenges when time series models are applied in marketing. 
Recently, \cite{Dekimpe2010} noted that time series modeling has just started to gain popularity in the marketing science community due to recent developments in marketing databases and computational power: \cite{Fader2010} used negative beta-geometric and beta-Bernoulli distributions with two latent variables (dead or alive) to model recursive discrete-time donation behavior, and recently, \cite{Lee2014} adopted a state space model to measure brand inertia.
%

In this paper, we present a new state space model for direct marketing, namely Dynamic Propensity Model (DPM).
The proposed model is a latent variable time series model that utilizes both marketing and purchase histories of a customer where the latent variable (state) represents a member's propensity to buy a product.
Our model tracks an individual's marketing touches and responses, then estimates his personalized probabilities for the first purchase at daily resolution.
Marketing histories contain the records of when and how many emails, direct mails, phone calls, display ads, and referral are sent or clicked. 
In the model, such marketing touches increase an individual's propensity, and this propensity score attenuates and propagates over time. 
We use six months' marketing records from one of the largest insurance companies in the U.S.
The degrees of the marketing effects and propagation strengths are statistically estimated from the data.
To estimate the parameters of the model,  a new statistical methodology has been developed, which combines particle methods and a stochastic gradient descent approach.

The design of DPM was, in fact, motivated by several findings.
Initially, our research goal was to find the retrospective correlation between marketing touches and purchase activities using generalized linear models.
During the investigation, however, we have observed that some of the marketing touches have \textit{negative effects} on purchase behaviors (see Section~\ref{sec:empirical}).
For some marketing practices, it is possible that some customers may be annoyed by frequent marketing touches.
Moreover, there are several aspects of the data that couldn't be addressed by a simple generalized model:
\begin{itemize}
\item Direct mails, emails, promotion phone calls are sent out to a strategically filtered group of customers. The company targeted persuadable customers, whose actuarial purchase probabilities are expected to be positively impacted by marketing touches.
\item Control and test groups are not readily identifiable for semi-targetable marketing events such as display ads and referrals. Uplift modeling cannot directly measure the effects of such semi-targetable marketing touches.
\item Past marketing touches have some effect on purchase activities. Appropriate effect time windows need to be determined to accurately estimate the true effects of marketing touches.
\end{itemize}
To address these findings, we needed a model that can track individual customers with different marketing histories.
The model also needed to provide consistent interpretations for building marketing strategies and monitoring customers' needs.
We view a purchase as an instantiation of marketing satisfaction \citep{Westbrook1987}, rather than a cognitive decision making process involving the semantic meaning of product attributes \citep{Oliver1980}. 
A dynamic process of satisfaction \citep{Fournier1999} naturally leads to algebraic linkages between temporally close satisfaction variables.
These findings and objectives led to the construction of our Dynamic Propensity Model.

Table~\ref{tab:notation} shows the mathematical notation used in this paper. The rest of this paper is organized as follows: 
In Section~\ref{sec:background}, we cover the basics of particle methods, parameter estimation techniques in state-space models, and stochastic gradient descent.
In Section~\ref{sec:dpm}, we formally introduce Dynamic Propensity model, and investigate the properties of the model.
The parameter estimation method for the proposed model is provided In Section~\ref{sec:estimation}, and
empirical results from the real-life dataset are illustrated in Section~\ref{sec:empirical}.
Finally, we discuss the limitations of the proposed method and future work in Section~\ref{sec:conclusion}.

\begin{table}\caption{Notation.}\label{tab:notation}
\begin{center}
\begin{tabular}{ l l  }
\hline
Symbol & Explanation \\
\hline\hline
$y_t^i$ & purchase indicator (output) \\
$\mathbf{r}_t^i$ & integer vector of non-actionable marketing touches (input)\\
$\mathbf{m}_t^i$ & integer vector of actionable marketing touches (input)\\ \hline
$x_t^i$ &  filtered propensity  (latent state)\\
$s_t^i$ & predictive propensity (auxiliary latent state)\\
$\varepsilon_t^i$ & standard normal noise i.e. $\varepsilon_t^i \sim \text{N}(0, 1)$\\\hline
$c$ & offset parameter for propensity  \\
$\phi$ & damping factor for propensity  \\
$\boldsymbol{\alpha}$ & coefficient vector for semi-targetable marketing touches e.g. display ads\\
$\boldsymbol{\beta}$ & coefficient vector for targetable marketing touches e.g. direct mail\\
$\boldsymbol{\theta}$ &  parameter vector i.e. $\boldsymbol{\theta}=(c, \phi, \boldsymbol{\alpha},\boldsymbol{\beta})$\\\hline
$\text{Pr}$ &  probability measure \\
$\mathcal{L}$ &  loss function or objective function \\
\hline
\end{tabular}
\end{center}
\end{table}

\section{Background}\label{sec:background}

In this section, we cover related work on particle methods, parameter estimation techniques in state-space models, and basics of stochastic gradient descent. These techniques form the building blocks of our proposed model, DPM, which is a latent variable time series model for large-scale enterprise size data. 
We found that traditional parameter estimation approaches for SSMs become intractable when applied to the size of our dataset. 
As a result, we developed a new parameter estimation technique by adopting (1) Monte Carlo simulation methods, and (2) stochastic optimization algorithms for the DPM objective function.

\subsection{Particle Methods}

If both the observation and the latent variables are normally distributed, the optimal filtering is known to be the Kalman Filter (KF) \citep{Kalman1960}.
For non-linear systems, several approximation techniques based on linearization, such as Extended KF (first-order approximation) and Unscented KF (second-order approximation), can be applied.
However, such linearization usually causes non-diminishing bias, and even worse, the algorithms are typically difficult to implement and tune correctly \citep{Julier2004}.  
Particle methods \citep{Gordon1993} use a different kind of approximation technique: Monte Carlo simulation.
Unlike those variants of KF, the state estimates from particle methods can be made arbitrarily accurate with enough particles.
Particle methods are based on a sequence of importance sampling steps.
Resampling techniques \citep{Liu1998} are typically adopted to decelerate the degeneracy of particles.  
Also, Auxiliary Particle Filter has been developed to prevent the degeneracy of the Sequential Monte Carlo \citep{Pitt1999}.
Particle methods are powerful general state-space estimation techniques that are widely applicable to non-linear evolution and observation processes \citep{Doucet:2008us}.
In this paper, we use a particle filter with resampling to estimate a sequence of dynamic propensity scores.

\subsection{Latent Variable Time Series Models}\label{sec:ssmest}

Latent variable time series models are categorized into two classes of models: state-space models (continuous latent variable) and hidden Markov models (discrete latent variable). 
Latent variables are effective for summarizing past observations, capturing an underlying dynamics, and providing human-interpretable results from complex observations \citep{JHo2013}.
\cite{raghavan2012} developed a hidden Markov model for modeling activity profiles of terrorist activities. 
\cite{xing2008} developed a state space model to capture time altering networks. 
\cite{valentini2013} employed a spatially structured factor analysis to model house prices, while \cite{nagaraja2011} used an autoregressive technique to capture the time effect. 
\cite{aktekin2013} used a Bayesian state space model to estimate mortgage default risk.

In this paper, we use a state space model to capture the (unobserved) propensity\footnote{Discrete latent variables in hidden Markov models cannot model continuous propensity scores.}.
Parameter estimation techniques for SSMs fall into three main groups: Bayesian online, maximum-likelihood offline, and maximum-likelihood online settings \citep{Kantas2009}.
In the Bayesian online setting, model parameters are assumed to be \textit{dynamic} over time series, and they are sequentially estimated.
Some of the successful algorithms are Liu-West Filter \citep{Liu2001}, Storvik Filter \citep{Storvik2002}, and Particle Learning \citep{Carvalho2010}.
For the maximum-likelihood online setting, \cite{Andrieu2005} have demonstrated an online estimation algorithm using block time series and pseudo-likelihood. Recall that the parameters in this paper are fixed but unknown.
 
In the offline (or batch) maximum-likelihood setting, two approaches have been popular: Fisher's scoring and Expectation-Maximization (EM).
The Fisher's scoring algorithm is a variant of Newton-Raphson algorithm based on the log-likelihood function.
However, obtaining the log-likelihood of a time series is typically intractable.
\cite{Doucet2003} proposed a general approach for approximating the log-likelihood using particle methods.
Although this Fisher's scoring algorithm is generally applicable to several settings, it is difficult to scale the gradients for high dimensional parameters \citep{Kantas2009}.
The EM algorithm is numerically more stable and usually computationally cheaper for high dimensional parameters. 
For a Gaussian SSM, the EM algorithm can be implemented using Kalman Filter and Smoother \citep{Shumway1982}.
For non-linear systems, \cite{Zia2008} introduced the EM-PF (EM using Particle Filter) algorithm, but many of the assumptions are not applicable in our setting.
For categorical time series, \cite{Park2014} proposed an efficient parameter estimation algorithm based on a stochastic EM algorithm \citep{Nielsen2000}.
However, their problem setting is slightly different from our setting; they assume different latent dynamics for individuals, and reasonably balanced class labels.
In this paper, we use an alternating maximization approach that is similar to the EM algorithm. 
There are some differences between our approach and the EM algorithm, which will be explained in Section~\ref{sec:estimation}.

\subsection{Stochastic Gradient Descent}

Stochastic Gradient Descent (SGD) is a large-scale optimization technique for a summation of differentiable objective functions. 
Consider an objective function $\mathcal{L}(\boldsymbol{\theta})$ based on a parameter $\boldsymbol{\theta}$ where the objective function can be written as $\mathcal{L}(\boldsymbol{\theta}) = \sum_{i } \mathcal{L}_i (\boldsymbol{\theta})$.
An iterative method like gradient descent can be used to reach a local optimum in expectation. If $\boldsymbol{\theta}^{(v)}$ is the estimate for the parameter in the $v$th  iteration, in the next iteration, a gradient descent algorithm produces:
\begin{align*}
\boldsymbol{\theta}^{(v+1)} = \boldsymbol{\theta}^{(v)} - \gamma^{(v)} \nabla \mathcal{L}(\boldsymbol{\theta}^{(v)})
= \boldsymbol{\theta}^{(v)} - \gamma^{(v)} \sum_i \nabla \mathcal{L}_i(\boldsymbol{\theta}^{(v)})
\end{align*}
where $\gamma^{(v)}$ is a suitable step size. The expectation $E(\mathcal{L}(\boldsymbol{\theta}))$ is to be calculated over the entire dataset for every update of $\boldsymbol{\theta}$. This can be costly for larger datasets. 

A \emph{stochastic} version of the gradient descent, SGD, is a more practical approach for large-scale learning problems \citep{Bottou2007}. 
SGD guarantees convergence to the optimal $\boldsymbol{\theta}$ while doing away with the costly expectation calculation \citep{Bottou1998}. 
The update step for SGD is simpler:
\begin{align*}
\boldsymbol{\theta}^{(v+1)} = \boldsymbol{\theta}^{(v)} - \gamma^{(v)} \sum_{i \in \mathcal{I}}\mathcal{L}_i (\boldsymbol{\theta}^{(v)})
\end{align*}
where $\mathcal{I}$ is a subset of the dataset. 
The subset can be as small as a single data point, though usually using small batches decreases the variance and leads to quicker convergence.
In this paper, we use SGD in our alternating maximization approach.

\section{Dynamic Propensity Model}\label{sec:dpm}

Dynamic Propensity Model (DPM) is a latent variable time series model that utilizes both marketing and purchase histories of a customer. 
The latent variable of DPM represents a member's propensity to buy a product.
The model tracks an individual' marketing touches and responses, and estimates his personalized probabilities for the first purchase at daily resolution.
These probability scores can be used in multiple applications: (1) predicting when the customer is likely to buy the product (within-customer application) and (2) targeting likely-to-buy customers (across-customer application).

\begin{figure}
\center
\includegraphics[width=0.8\textwidth]{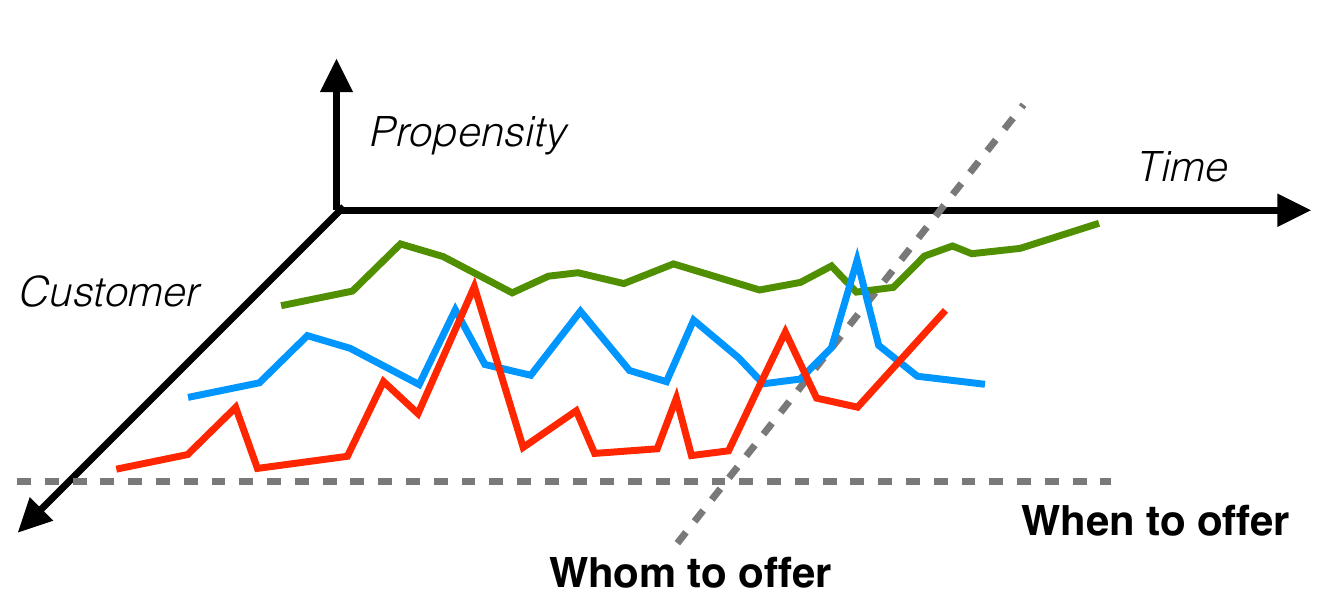}
\caption{The connection between \textit{whom} to offer and \textit{when} to offer.}\label{fig:duality}
\end{figure}

DPM uses a different modeling strategy from traditional targeting models.
Traditional targeting models estimate cross-sectional probabilities, whereas DPM models longitudinal probabilities.
In other words, traditional targeting models answer whom to offer, and DPM suggests when to offer.
Although these two goals may seem fundamentally different, one indirectly indicates the other given a limited marketing budget.
This connection is shown in Figure~\ref{fig:duality}.
In cross-sectional modeling, choosing a subset of customers is typically based on the rank orders of probability scores at a given time.
For a different cross-section of data, a different subset of customers will be chosen.
As a result, a customer may or may not be chosen for direct marketing at a given time, and this property indirectly determines when to offer.

Although their outcomes may be superficially similar, each approach has different modeling limitations.
Cross-sectional models are easy to estimate, but complex effects, such as time-varying effects and customer heterogeneity, are difficult to model.
On the other hand, longitudinal models can capture dynamic and heterogeneous effects, but they typically need more samples for parameter estimation, and sometimes, some model parameters are almost intractable to estimate.
Due to the complexity of time-series models, cross-sectional models are traditionally modified to handle complex effects, instead of directly using time-series models.
\cite{Allenby1994} applied temporally correlated error terms for modeling household purchase behavior.
\cite{Keane1997} suggested a variant of discrete choice model that captures both customer heterogeneity and temporal dependency.

Our goal is to build a time series model that can easily capture customer heterogeneity and temporal dependency, and furthermore, the model should be simple enough to extend with other covariates.
In this paper, we demonstrate DPM on one product and associated marketing touches for promoting that product, although it can be extended to deal with multiple products through a matrix representation. 
In other words, marketing touches for promoting different products are ignored to keep the simplicity of our illustration.
Recall that, in DPM,  a customer has a latent factor (propensity) that evolves over time.
We assume that marketing effects can be super-positioned i.e. they additively affect the propensity.

A direct application of state-space model can capture these listed properties:
\begin{align*}
\text{(Filtered Propensity)} &\quad x_{t+1}^i = c_d + \phi_d x_{t}^i + \boldsymbol{\alpha}^\top_d \mathbf{r}_t^i + \boldsymbol{\beta}^\top_d \mathbf{m}_{t}^i +  \varepsilon_{t}^i \\
\text{(Tomorrow's Purchase)} &\quad \text{Pr}(y_{t+1}^i = 1) = \frac{1}{1+\exp(-x_{t+1}^i)}
\end{align*}
where the superscript $i$ represents the $i$th customer, and the subscript $t$ indicates variables at time $t$.
The subscript $d$ on the model parameters indicates a demographic segment that exhibit similar responses to marketing touches i.e. homogeneous marketing response group.
Two different types of marketing touches are illustrated: semi-targetable $\mathbf{r}_t^i$ and targetable $\mathbf{m}_t^i$ marketing touches. 
The targetable marketing touches include emails, direct mails, and phone calls. In these cases, marketing can be directly targeted to a particular customers.
The semi-targetable marketing touches represent display ads and referrals.
Although a company can control the exposure to such semi-targetable marketing touches, such marketing touches cannot be targeted to a specific individual and the exposure involves a certain degree of randomness.
Some semi-targetable marketing variables contain marketing responses e.g. clicking display ads, and we view that these activities increase the propensity.

This direct application of state space model, however, faces two practical issues as follows:
\begin{itemize}
\item \textbf{Time difference within a day}: Even though $\mathbf{m}_t$ and $y_t$ are indexed by the same subscript, one occurs before the other e.g. $\mathbf{m}_t$ in the morning and $y_t$ in the afternoon. In this paper, we assume that a purchase decision $y_t$ is made at the very start of a day, thus $\mathbf{m}_t$ and $\mathbf{r}_t$ affect the next day's purchase decision $y_{t+1}$.
\item \textbf{No data anchor on marketing effects}: If $\mathbf{x}$ is known, then estimating $\boldsymbol{\alpha}$ and $\boldsymbol{\beta}$ only depends on $\mathbf{x}$ in the na\"{i}ve SSM application. If the estimates of $\mathbf{x}$ is noisy, then the error would propagate to the parameter estimates. In fact, we have observed that parameters do not converge when using this model. The estimated parameters should be anchored on actual data samples, rather than being determined by latent variables.
\end{itemize}
These issues can be addressed by introducing an auxiliary variable $s_{t+1}$ between $x_{t}$ and $x_{t+1}$.

\begin{figure}[h]
\centering
\definecolor{mycolor1}{RGB}{217,91,67}
\definecolor{mycolor2}{RGB}{192,41,66}
\definecolor{mycolor3}{RGB}{84,36,55}
\definecolor{mycolor4}{RGB}{83,119,122}
\tikzstyle{state}=[circle,thick,minimum size=1.2cm,draw=mycolor1]
\tikzstyle{measurement}=[circle,thick,minimum size=1.2cm,draw=mycolor2]
\tikzstyle{input}=[circle,thick,minimum size=1.2cm,draw=mycolor3]
\tikzstyle{matrx}=[rectangle,thick,minimum size=1cm,draw=mycolor4]
\tikzstyle{background}=[rectangle,fill=gray!10,inner sep=0.2cm,rounded corners=5mm]
\begin{subfigure}[b]{0.8\textwidth}
\begin{tikzpicture}[>=latex,text height=1.5ex,text depth=0.25ex]
  \matrix[row sep=0.5cm,column sep=0.3cm] {
        \node (m_t) [input] {$\mathbf{m}_{t}$};  & 
        \node (r_t)   [input] {$\mathbf{r}_{t}$};   &
        \node (m_t+1)   [input] {$\mathbf{m}_{t+1}$};  & 
        \node (r_t+1)   [input] {$\mathbf{r}_{t+1}$}; &
        \node (m_t+2)   [input] {$\mathbf{m}_{t+2}$};  & 
        \node (r_t+2)   [input] {$\mathbf{r}_{t+2}$}; &
        \\
        \node (s_t-1) {$\cdots$};           &
        \node (x_t) [state] {${x}_{t}$}; & &
        \node (x_t+1) [state] {${x}_{t+1}$}; & &         
        \node (x_t+2) [state] {${x}_{t+2}$}; &
        \node (s_t+2)         {$\cdots$};                   
        \\
        &\node (y_t)   [measurement] {${y}_{t}$}; &     
        & \node (y_t+1) [measurement] {${y}_{t+1}$}; & 
        & \node (y_t+2) [measurement] {${y}_{t+2}$}; &         
        \\
    };    
    \path[->]
        (s_t-1) edge[thick] node[above] {$\boldsymbol{\theta}$} (x_t)	
        (x_t)   edge[thick] node[above] {$\boldsymbol{\theta}$} (x_t+1)		
        (x_t+1) edge[thick] node[above] {$\boldsymbol{\theta}$} (x_t+2)
       (x_t+2) edge[thick] node[above] {$\boldsymbol{\theta}$} (s_t+2)        
        
        (x_t) edge (y_t)				
        (x_t+1)   edge (y_t+1)
         (x_t+2)   edge (y_t+2)       
       
        (r_t)   edge (x_t)        
        (m_t)   edge (x_t)
        (r_t+1)   edge (x_t+1)
        (m_t+1)   edge (x_t+1)
        (r_t+2)   edge (x_t+2)
        (m_t+2)   edge (x_t+2)        
        ;
\end{tikzpicture}
\caption{Na\"{i}ve Application of SSM}
\end{subfigure}

\begin{subfigure}[b]{0.8\textwidth}
\begin{tikzpicture}[>=latex,text height=1.5ex,text depth=0.25ex]
  \matrix[row sep=0.5cm,column sep=0.3cm] {
        \node (m_t-1)   [input] {$\mathbf{m}_{t-1}$};   &
        \node (r_t-1)   [input] {$\mathbf{r}_{t-1}$};     & 
        \node (m_t)   [input] {$\mathbf{m}_{t}$};  &
        \node (r_t) [input] {$\mathbf{r}_{t}$}; & 
        \node (m_t+1)   [input] {$\mathbf{m}_{t+1}$};  &
        \node (r_t+1) [input] {$\mathbf{r}_{t+1}$}; & &
        \\
        \node (x_t-1) {$\cdots$}; &
        \node (s_t) [matrx] {$s_{t}$};       &
        \node (x_t)   [state] {${x}_t$};     &
        \node (s_t+1)   [matrx] {$s_{t+1}$};       &
        \node (x_t+1) [state] {${x}_{t+1}$}; &
        \node (s_t+2)  [matrx] {$s_{t+2}$};       &      
        \node (x_t+2) [state] {${x}_{t+2}$}; &
        \node (s_t+3)         {$\cdots$};           
        \\
	& &
        \node (y_t)   [measurement] {${y}_{t}$};     &
        &
        \node (y_t+1) [measurement] {${y}_{t+1}$}; &
        &
        \node (y_t+2) [measurement] {${y}_{t+2}$}; &        
        \\
    };    
    \path[->]
        (x_t-1) edge[thick] node[above] {$\boldsymbol{\theta}$} (s_t)	
        (s_t) edge[thick] (x_t)		
        (x_t)   edge[thick] node[above] {$\boldsymbol{\theta}$} (s_t+1)		
        (s_t+1)   edge[thick] (x_t+1)	
        (x_t+1) edge[thick] node[above] {$\boldsymbol{\theta}$} (s_t+2)
        (s_t+2)   edge[thick] (x_t+2)	
        (x_t+2) edge[thick] node[above] {$\boldsymbol{\theta}$} (s_t+3)
                
        (s_t) edge (y_t)				
        (s_t+1)   edge (y_t+1)
        (s_t+2)   edge (y_t+2)
                
        (m_t-1)   edge (s_t)
        (r_t-1)   edge (s_t)
        (m_t)   edge (s_t+1)
        (r_t)   edge (s_t+1)                
        (m_t+1)   edge (s_t+2)
        (r_t+1)   edge (s_t+2)                        
        ;
\end{tikzpicture}
\caption{Dynamic Propensity Model}
\end{subfigure}
\caption{Graphical Models for the simple SSM and DPM. The path from $x_t$ to $y_{t+1}$ in DPM involves the model parameter $\boldsymbol{\theta}$, while the SSM path from $x_t$ to $y_{t+1}$ is blocked by $x_{t+1}$. }\label{fig:graphical}
\end{figure}
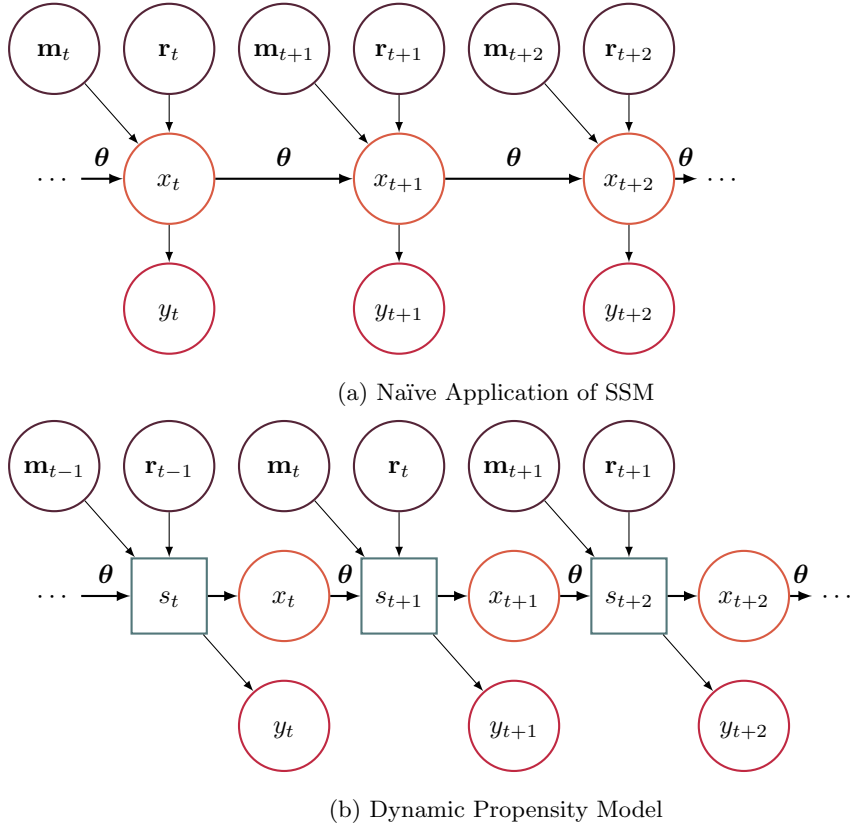

We now introduce the model equations for DPM:
\begin{align*}
\text{(Propagated Propensity)} &\quad\quad x_{t}^i = s_t^i + \varepsilon_{t}^i\\
\text{(Predicted Propensity)} &\quad\quad s_{t+1}^i = c_d + \phi_d x_{t}^i + \boldsymbol{\alpha}^\top_d \mathbf{r}_t^i + \boldsymbol{\beta}^\top_d \mathbf{m}_{t}^i  \\
\text{(Tomorrow's Purchase)} &\quad\quad \text{Pr}(y_{t+1}^i = 1) = \frac{1}{1+\exp(-s_{t+1}^i)}
\end{align*}
where $\varepsilon_t^i$ is drawn from the standard normal distribution. 
The auxiliary variable $s_{t+1}$ is deterministic given today's propensity $x_{t}$ and marketing touches $\mathbf{m}_t$ and $\mathbf{r}_t$.
This auxiliary variable can be viewed as the propensity in the evening that directly affects the purchase decision at the very start of the next day.
Figure~\ref{fig:graphical} shows and compares the graphical representations of the naive SSM application and DPM.

The auxiliary variable in DPM not only refines the time resolution of purchase process, but also provides a new perspective on parameter estimation. 
As an illustrative example, let us consider an alternating maximization approach for parameter estimation. 
In the SSM formulation, we basically solve two maximization problems:
\begin{equation*}
\begin{aligned}[c]
\max~ \boldsymbol{\theta} &\mid \mathbf{x}, \mathbf{y} \\
\max~ \mathbf{x} &\mid \boldsymbol{\theta}, \mathbf{y}
\end{aligned}\quad\Longrightarrow\quad
\begin{aligned}[c]
\max~ \boldsymbol{\theta} &\mid \mathbf{x} \\
\max~ \mathbf{x} &\mid \boldsymbol{\theta}, \mathbf{y}
\end{aligned}
\end{equation*}
where $\boldsymbol{\theta} = \{c, \phi, \boldsymbol{\alpha}, \boldsymbol{\beta}\}$.
Note that $\mathbf{y}$ in the first maximization problem is removed, since $\boldsymbol{\theta}$ is solely determined by $\mathbf{x}$ i.e. data-anchor issue.
On the other hand, in DPM, we solve three maximization problems with respect to $\boldsymbol{\theta}$, $\mathbf{x}$, and $\mathbf{s}$:
\begin{equation*}
\begin{aligned}[c]
\max~ \boldsymbol{\theta} &\mid \mathbf{x}, \mathbf{y}, \mathbf{s} \\
\max~ \mathbf{x} &\mid \boldsymbol{\theta}, \mathbf{y}, \mathbf{s}\\
\max~  \mathbf{s} &\mid \boldsymbol{\theta}, \mathbf{x}, \mathbf{y} 
\end{aligned}\quad\Longrightarrow\quad
\begin{aligned}[c]
\max~\boldsymbol{\theta} &\mid \mathbf{x}, \mathbf{y} \\
\max~ \mathbf{x} &\mid \boldsymbol{\theta}, \mathbf{y}\\
(  \mathbf{s} &\mid \boldsymbol{\theta}, \mathbf{x}, \mathbf{y} )
\end{aligned}
\end{equation*}
where the last maximization problem is trivial as $\mathbf{s}$ is determined by $\mathbf{x}$ and $\boldsymbol{\theta}$.
In fact, these three maximization problems are actually two maximization problems. 
To remove the last maximization problem, we treat the auxiliary variable as a nuisance variable.
We remove $\mathbf{s}$ using the relation $s_{t+1}=c_d + \phi_d x_{t} + \boldsymbol{\alpha}^\top_d \mathbf{r}_t + \boldsymbol{\beta}^\top_d \mathbf{m}_{t} $.
Both the resultant two maximization problems are now anchored by the observation $\mathbf{y}$.
The path from $x_t$ to $y_{t+1}$ in DPM involves the model parameter $\boldsymbol{\theta}$ (see Figure~\ref{fig:graphical}).
The auxiliary variable in DPM resulted in a different set of maximization problems from the one from the SSM formulation.

The complete joint probability distribution of DPM for a particular customer can be recursively factorized as follows:
\begin{align*}
&\text{Pr}(\mathbf{R}, \mathbf{M}, \mathbf{y}, \mathbf{x}) \\
&= \underbrace{\text{Pr}(\mathbf{R}, \mathbf{M} )}_{\text{marketing strategy}} \prod_t  \text{Pr}(y_{t+1} \mid \underbrace{x_t}_{\mathclap{\text{customer heterogeneity}}},\mathbf{r}_t, \mathbf{m}_t ) \underbrace{\text{Pr}(x_t \mid x_{t-1} \mathbf{r}_{t-1}, \mathbf{m}_{t-1} )}_{\text{temporal dependency}}
\end{align*}
where we removed the superscript $i$ to make notation simple.
As can be seen, the model captures both customer heterogeneity and temporal dependency.
Note that, in DPM, customer heterogeneity originates from differences in marketing touches and responses over time. 

In fact, customer heterogeneity can be captured by demographic segments such as age, income, and family size.
This demographic heterogeneity can affect customers' base purchase rates and response dynamics.
Note that, in our DPM formulation, the model parameters are indexed by the demographic segment variable $d$.
In practice, we build a separate model for each demographic segment that exhibits homogeneous behavioural patterns.
For clarity, we focus on one demographic segment from here onwards, dropping the index $d$.

\section{Parameter Estimation}\label{sec:estimation}

The maximum likelihood parameter for DPM is hard to optimize, as can be seen from the likelihood equation:
\begin{align*}
&\arg\max_{\boldsymbol{\theta}} \prod_{i=1}^N \prod_{t=1}^{T^i} \text{Pr}_{\boldsymbol{\theta}}(y_{t+1}^i \mid \mathbf{r}_t^i, \mathbf{m}_t^i) 
\end{align*}
where $N$ is the total number of customers, and $T^i$ is the number of days that the $i$th customer is monitored.
The likelihood is defined on the observed variables: $y_t^i$, $\mathbf{m}_t^i$, and $\mathbf{r}_t^i$.
To obtain the actual value of the likelihood, we need to integrate out the latent states:
\begin{align*}
&\arg\max_{\boldsymbol{\theta}} \prod_{i=1}^N \int_{\mathbf{x}} \prod_{t=1}^{T^i} \text{Pr}_{\boldsymbol{\theta}}(y_{t+1}^i \mid {x}_t^i, \mathbf{r}_t^i, \mathbf{m}_t^i)\text{Pr}_{\boldsymbol{\theta}}({x}_t^i \mid {x}_{t-1}^i \mathbf{r}_{t-1}^i, \mathbf{m}_{t-1}^i) d\mathbf{x}\\
& = \arg\max_{\boldsymbol{\theta}} \sum_{i=1}^N  \log \int_{\mathbf{x}} \prod_{t=1}^{T^i} \text{Pr}_{\boldsymbol{\theta}}(y_{t+1}^i \mid {x}_t^i, \mathbf{r}_t^i, \mathbf{m}_t^i)\text{Pr}_{\boldsymbol{\theta}}({x}_t^i \mid {x}_{t-1}^i \mathbf{r}_{t-1}^i, \mathbf{m}_{t-1}^i)  d\mathbf{x} 
\end{align*}
where $\mathbf{x}$ represents $x_{1:T^i}^i$. The inner integral makes optimization of the log-likelihood intractable.
This type of maximization problem has been traditionally approached by the EM algorithm \citep{Neal1998}. 
The EM algorithm derives a lower bound of the log-likelihood, and then maximizes this lower bound.
The lower bound is obtained using Jensen's inequality for any arbitrary distribution $\text{Q}^i(\mathbf{x})$ :
\begin{align}
 \sum_{i=1}^N  \int_{\mathbf{x}} \text{Q}^i(\mathbf{x}) \log  \frac{\prod_{t=1}^{T^i} \text{Pr}_{\boldsymbol{\theta}}(y_{t+1}^i \mid {x}_t^i, \mathbf{r}_t^i, \mathbf{m}_t^i)\text{Pr}_{\boldsymbol{\theta}}({x}_t^i \mid {x}_{t-1}^i \mathbf{r}_{t-1}^i, \mathbf{m}_{t-1}^i) }{\text{Q}^i(\mathbf{x})} d\mathbf{x} \label{eq:em}
\end{align}

This lower bound is maximized when
\begin{align*}
\text{Q}^i(\mathbf{x}) = \prod_{t=1}^{T^i} \text{Pr}_{\boldsymbol{\theta}}(y_{t+1}^i \mid {x}_t^i, \mathbf{r}_t^i, \mathbf{m}_t^i)\text{Pr}_{\boldsymbol{\theta}}({x}_t^i \mid {x}_{t-1}^i \mathbf{r}_{t-1}^i, \mathbf{m}_{t-1}^i)
\end{align*}
For dynamic linear systems with Gaussian noise, $\text{Q}^i(\mathbf{x})$ can be obtained in a closed form.
However, the binary purchase indicators cannot be directly modeled as numeric variables, and the form of $\text{Q}^i(\mathbf{x})$ should be numerically approximated.
The EM algorithm for DPM can be written as follows:
\begin{enumerate}
\item Initialize $\boldsymbol{\theta}$
\item Until convergence,
	\begin{enumerate}
	\item (E-step) Estimate $Q^i$ that maximizes Equation~\ref{eq:em}
	\item (M-step) Estimate $\boldsymbol{\theta}$ that maximizes Equation~\ref{eq:em}
	\end{enumerate}
\end{enumerate}
This method is practical only for a small $N$, but not for the size of our data. 
If we use particle filters for the E-step, the number of simulation particles to store is $P \times  \sum_{i} T^i $.
As an illustrative example, if we use 5000 particles for 200K customers with average 100 observations per customer, a hundred million particles need to be stored and flushed at every iteration.
Furthermore, the M-step needs to run on the massive size of particles, which is also a challenging engineering problem.

To facilitate practical optimization, we optimize a surrogate instead of the true likelihood function by viewing $x_t$ as a temporally correlated random offset parameter, rather than a latent variable.
This view helps us maximize the \textit{conditional likelihood} function of DPM, instead of the marginal likelihood.
Our goal is to obtain estimates for both $\mathbf{X}=\{\mathbf{x}^1, \mathbf{x}^2, \ldots, \mathbf{x}^N\}$ and $\boldsymbol{\theta}$.
The original optimization problem with the parameter set $\boldsymbol{\theta}$ now transforms to a new optimization problem with two sets of variables as follows: 
\begin{align}
\label{eqn:LDPM}
\begin{split}
&\arg\max_{\boldsymbol{\theta},\mathbf{X}} \mathcal{L}_{\text{DPM}} (\boldsymbol{\theta},\mathbf{X}) \\
& = \arg\max_{\boldsymbol{\theta},\mathbf{X}} \prod_{i=1}^N \prod_{t=1}^{T^i} \text{Pr}_{\boldsymbol{\theta}}(y_{t+1}^i \mid {x}_{t}^i,\mathbf{r}_t^i, \mathbf{m}_t^i )\text{Pr}_{\boldsymbol{\theta}}({x}_t^i \mid {x}_{t-1}^i, \mathbf{r}_{t-1}^i, \mathbf{m}_{t-1}^i) \\
& = \arg\max_{\boldsymbol{\theta},\mathbf{X}} \sum_{i=1}^N \sum_{t=1}^{T^i} 
\underbrace{\log \text{Pr}_{\boldsymbol{\theta}}(y_{t+1}^i \mid {x}_{t}^i,\mathbf{r}_{t}^i, \mathbf{m}_t^i )}_{\text{generalized linear model}} + 
\underbrace{\log \text{Pr}_{\boldsymbol{\theta}}({x}_t^i \mid {x}_{t-1}^i, \mathbf{r}_{t-1}^i,\mathbf{m}_{t-1}^i)}_{\text{time series}} 
\end{split}
\end{align}

where $\mathcal{L}_{\text{DPM}}$ is the objective function of DPM. 
As can be seen, the inner integral is removed since $\mathbf{X}$ is treated as a parameter matrix.
The factorized form in Equation~\ref{eqn:LDPM} provides insightful explanation about the objective function.
The first part of the factorized form is a logistic regression model with customer heterogeneity, while the second part explains temporal dependency.
This objective function can be also viewed as a logistic regression with sophisticated temporal regularization, or a time series model with a logistic loss regularization.

An alternating maximization of the objective function can be written as follows:
\begin{enumerate}
\item Initialize $\boldsymbol{\theta}$
\item Until convergence on $\boldsymbol{\theta}$,
	\begin{enumerate}
	\item Maximize $\mathcal{L}_{\text{DPM}}$ over $\mathbf{X}$ using Particle Methods
	\item Maximize $\mathcal{L}_{\text{DPM}}$ over $\boldsymbol{\theta}$
	\end{enumerate}
\end{enumerate}
This procedure is more efficient than the brute-force EM algorithm, since only the most likely single path $x_{1:T}$ needs to be stored.
However, the maximization over $\boldsymbol{\theta}$ can be still expensive because of large number of customer samples.
Stochastic Gradient Descent (SGD) is a suitable solution to this type of large-scale learning problem.
If we apply SGD on the second maximization step, we obtain:
\begin{enumerate}
\item Initialize $\boldsymbol{\theta}$
\item Until convergence on $\boldsymbol{\theta}$,
	\begin{enumerate}
	\item Maximize $\mathcal{L}_{\text{DPM}}$ over $\mathbf{X}$
	\item Initialize $\boldsymbol{\theta}_{\text{SGD}} = \boldsymbol{\theta}$
	\item Until convergence on $\boldsymbol{\theta}_{\text{SGD}}$,
		\begin{enumerate}
		\item Select a subset of customer samples,
			\begin{enumerate}
			\item Maximize $\mathcal{L}_{\text{DPM}}$ over $\boldsymbol{\theta}_{\text{SGD}}$
			\end{enumerate}
		\end{enumerate}
	\end{enumerate}
\end{enumerate}
where $\boldsymbol{\theta}_{\text{SGD}}$ represents the inner loop parameter in the SGD step.
This procedure is the same as the previous procedure except the SGD maximization part.
To estimate $\boldsymbol{\theta}$, a subset of customers are uniformly sampled from the entire training dataset.
The gradient of the objective function is calculated based on the subset, then the inner loop parameter is incremented by the gradient.
This inner loop continues until $\boldsymbol{\theta}_{\text{SGD}}$ converges.

The maximization step of $\mathbf{X}$ still requires a particle method that scans the entire training dataset.
Note that not all $\mathbf{x}^i$ samples are used in the SGD maximization step.
The particle estimates on $\mathbf{x}^i$ can be generated on demand according to the selected subset as follows:
\begin{enumerate}
\item Initialize $\boldsymbol{\theta}$
\item Until convergence on $\boldsymbol{\theta}$,
	\begin{enumerate}
	\item Initialize $\boldsymbol{\theta}_{\text{SGD}} = \boldsymbol{\theta}$
	\item Until convergence on $\boldsymbol{\theta}_{\text{SGD}}$,
		\begin{enumerate}
		\item Select a subset of samples,
			\begin{enumerate}
			\item Maximize $\mathcal{L}_{\text{DPM}}$ over $\mathbf{X}$ using $\boldsymbol{\theta}$
			\item Maximize $\mathcal{L}_{\text{DPM}}$ over $\boldsymbol{\theta}_{\text{SGD}}$
			\end{enumerate}
		\end{enumerate}
	\end{enumerate}
\end{enumerate}
Although $\mathbf{X}$ is estimated only on the selected subset of samples, this procedure has nested loops. 
The inner loop estimates the stochastic gradient parameter, and the outer loop resets the stochastic gradient parameter.
We observe that the nested loops can be effectively removed, since in practice the inner loop converges within one or two iterations for customer samples of size 1.

We now introduce the parameter estimation algorithm for DPM in Algorithm~\ref{algorithm:dpm}.
\begin{algorithm}
\SetAlgoLined
\KwData{$\mathbf{Y}$}
\KwResult{$\boldsymbol{\theta}$}
Initialize $\boldsymbol{\theta}$\;
	\While{Until convergence on $\boldsymbol{\theta}$}{
		Randomly pick a customer $\mathbf{y}^i = [y_1, y_2, \ldots, y_{T^i}]$\;
		$\boldsymbol{\theta} = \boldsymbol{\theta} + \gamma  \partial_{\boldsymbol{\theta}} \mathcal{L}_{\text{DPM}} (\boldsymbol{\theta}, \argmax_{\mathbf{x}^i} \mathcal{L}_{\text{DPM}} (\boldsymbol{\theta}, \mathbf{x}^i))$\;
	}
\caption{SGDPM: DPM parameter estimation algorithm using stochastic gradient}\label{algorithm:dpm}
\end{algorithm}

The SGDPM algorithm converges almost surely if $\sum_{v} (\gamma^{(v)})^2 < \infty$ and $\sum_{v} \gamma^{(v)} = \infty$.
Under sufficient regularity conditions, the best convergence speed is achieved when $\gamma^{(v)} \sim v^{-1}$ \citep{Murata1998}.
The algorithm converges to local optima, thus for potentially better solutions, one needs to try out several different initializations.
The inner $\argmax$ is estimated using particle methods, and the outer gradient can be obtained in a closed form.
As an illustrative example, we show a gradient form for the damping parameter $\phi$:
\begin{small}
\begin{align*}
& \partial_{\phi} \sum_{t=1}^{T^i}  \log \text{Pr}_{\boldsymbol{\theta}}(y_{t+1}^i \mid {x}_{t}^i,\mathbf{m}_t^i ) + \log \text{Pr}_{\boldsymbol{\theta}}({x}_t^i \mid {x}_{t-1}^i, \mathbf{m}_{t-1}^i)\\
& = \partial_{\phi} \sum_{t=1}^{T^i} \log (\frac{\exp(s_{t+1}^i )}{1+\exp(s_{t+1}^i )})^{y_{t+1}^i}(\frac{1}{1+\exp(s_{t+1}^i )})^{1-y_{t+1}^i} - \frac{1}{2}\log 2\pi \sigma^2  - \frac{(x_t^i - s_t^i )^2}{2\sigma^2}\\
& = \sum_{t=1}^{T^i} y_{t+1} x_{t}^i + \frac{x_{t}^i \exp(c + \phi x_{t}^i + \boldsymbol{\alpha}^\top \mathbf{r}_t^i + \boldsymbol{\beta}^\top \mathbf{m}_{t}^i)}{1+ \exp(c + \phi x_{t}^i + \boldsymbol{\alpha}^\top \mathbf{r}_t^i + \boldsymbol{\beta}^\top \mathbf{m}_{t}^i)} \\
&\quad\quad\quad\quad\quad + \frac{x_{t-1}^i (x_t^i - c - \phi x_{t-1}^i - \boldsymbol{\alpha}^\top \mathbf{r}_{t-1}^i - \boldsymbol{\beta}^\top \mathbf{m}_{t-1}^i)}{\sigma^2}
\end{align*}
\end{small}
The SGD update for the damping parameter $\phi$ is as follows:
\begin{align*}
{\phi}^{(v+1)} = {\phi}^{(v)} + \gamma^{(v)}  \partial_\phi \mathcal{L}_{\text{DPM}} (\boldsymbol{\theta}^{(v)}, \mathbf{x}^i )
\end{align*}
where the gradient is added, since we want to maximize the objective function.
 SGD updates for the other parameters can be similarly obtained, we omit detailed equations for conserving space.

\section{Empirical Study}\label{sec:empirical}

In this section, we evaluate DPM using a real dataset from one of the largest insurance companies in the U.S.
We first give an overview of the dataset and its basic statistics, and show evidence for the time dependency of marketing effects.
Baseline models are constructed using logistic regression models with lagged variables. 
Finally, we compare the parameters and predictive performances of DPM and the baseline models.

\subsection{Dataset Overview}


The dataset contains six months' records of marketing touches, responses, and purchases on 13 different kinds of financial products, collected over July 2012 - December 2012.
There are six different types of marketing touches: three semi-targetable $\mathbf{r}_t^i = (r_{1t}^i,r_{2t}^i,r_{3t}^i)$, three targetable $\mathbf{m}_t^i = (m_{1t}^i,m_{2t}^i,m_{3t}^i)$ marketing touches.
Marketing touch types are masked due to confidentiality reasons.
The records are samples from the company's customer database, stratified based on products and customers who have not yet bought the respective product in the beginning of our six months time window.
The company believes that, if personalized marketing touches can engage these new customers, then they tend to turn into loyal customers.
Among the 13 different product types, for illustrative purpose, DPM is applied to the first purchases of two products: Product A and Product B.
Customers are partitioned into training (50\%) and test datasets (50\%).

\begin{table}[h]
\caption{Data Format }\label{tab:format}
\begin{center}
\begin{tabular}{ l | l |  l l l l l l | l }
\hline
id & time  & $r.1$  & $r.2$ & $r.3$ & $m.1$  & $m.2$ & $m.3$ & y \\
\hline\hline
1847410 & July 1 & 0 & 0 & 0 & 1 & 0 & 0 & 0\\
1847410 & July 2 & 0 & 0 & 0 & 0 & 0 & 0 & 0\\
 & $\vdots$ &  &  &  &  &  &  & \\
1847410 & Dec 2 & 1 & 2 & 0 & 0 & 0 & 0 & 0\\ 
1847410 & Dec 3 & 1 & 0 & 0 & 0 & 0 & 0 & \textbf{1} (purchase) \\ \hline
1352132 & July 1 & 0 & 0 & 0 & 0 & 0 & 0 & 0\\ 
 & $\vdots$ &  &  &  &  &  &  & \\
\end{tabular}
\end{center}
\end{table}

\begin{table}[h]
\caption{Dataset Overview }\label{tab:data}
\begin{center}
\begin{tabular}{ l l l l l }
\hline
 & \multicolumn{2}{l }{Product A}  & \multicolumn{2}{l }{Product B}  \\
  & \multicolumn{2}{l }{(70K customers)}  & \multicolumn{2}{l }{(20K customers)}  \\
Variable & Mean  & Max  & Mean  & Max  \\
\hline\hline
$r_1$ & 0.0010 &  2 & 0.0115 & 3\\
$r_2$ & 0.0044 &  3 & 0.0057 & 2\\
$r_3$ & 0.0004 &  1 & 0.0027 & 2\\
$m_1$ & 0.0165 & 1 & 0.0168 & 1 \\
$m_2$ & 0.0354 & 1 & 0.0229 & 1\\
$m_3$ & 0.0003 &  1 & 0.0032 & 1 \\
  $y$ & 0.0001 &  1 & 0.0004 & 1 \\
\hline
\end{tabular}
\end{center}
\end{table}

The format of the data is shown in Table~\ref{tab:format}. 
Each row is identified by the combination of IDs and time stamps.
Marketing touches are recorded as counts i.e. $r.2=2$ means two events of the same marketing type for a day.
Table~\ref{tab:data} shows basic statistics of the data.
The dataset for Product A contains about 70,000 customers, and the dataset for Product B has about 20,000 customers.
Both the marketing touches and the targets are highly sparse.
For example, daily purchase rate of Product A is 0.01\%, and 0.04\% for Product B.

\begin{figure}
\center
\begin{subfigure}[b]{0.45\textwidth}
	\includegraphics[width=1\textwidth]{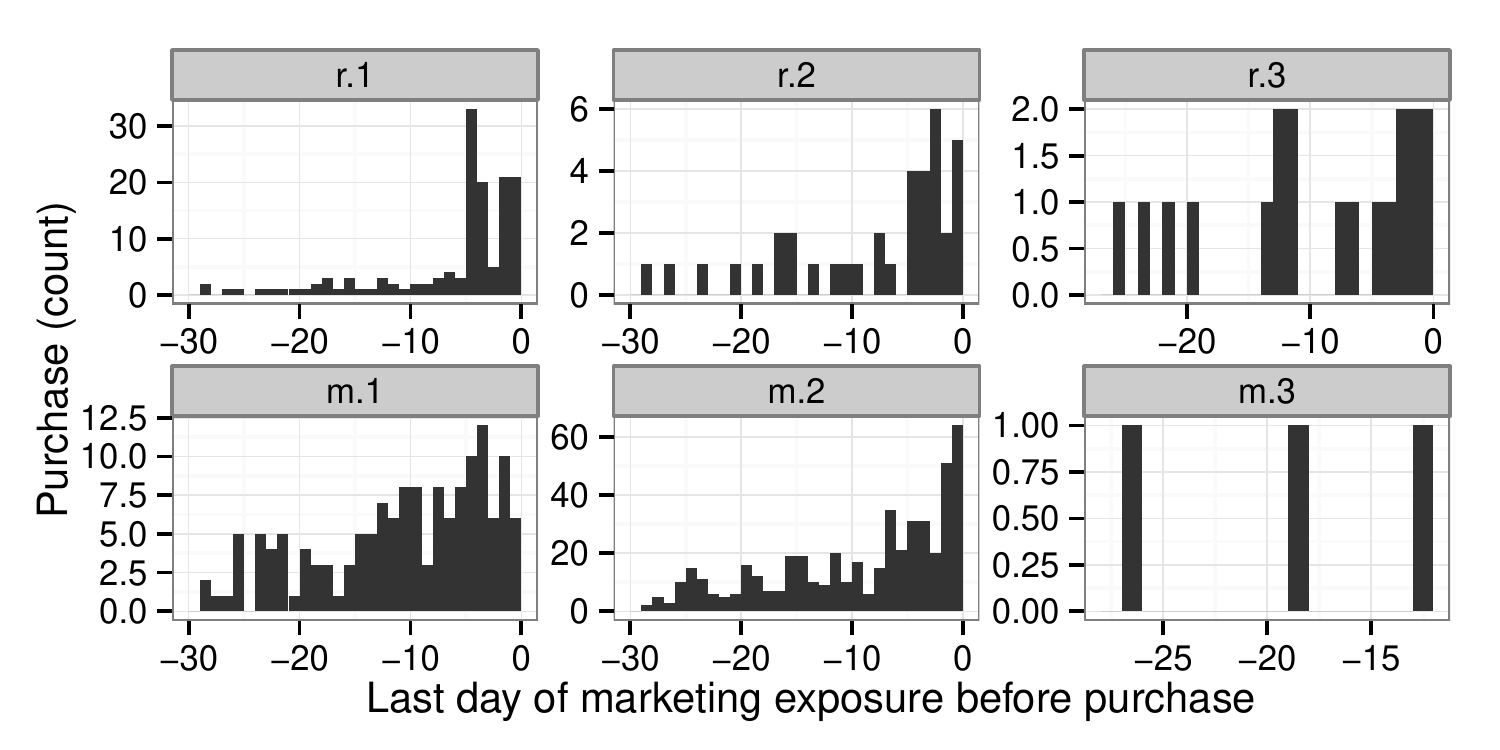}
	\caption{Product A}
\end{subfigure}
\begin{subfigure}[b]{0.45\textwidth}
	\includegraphics[width=1\textwidth]{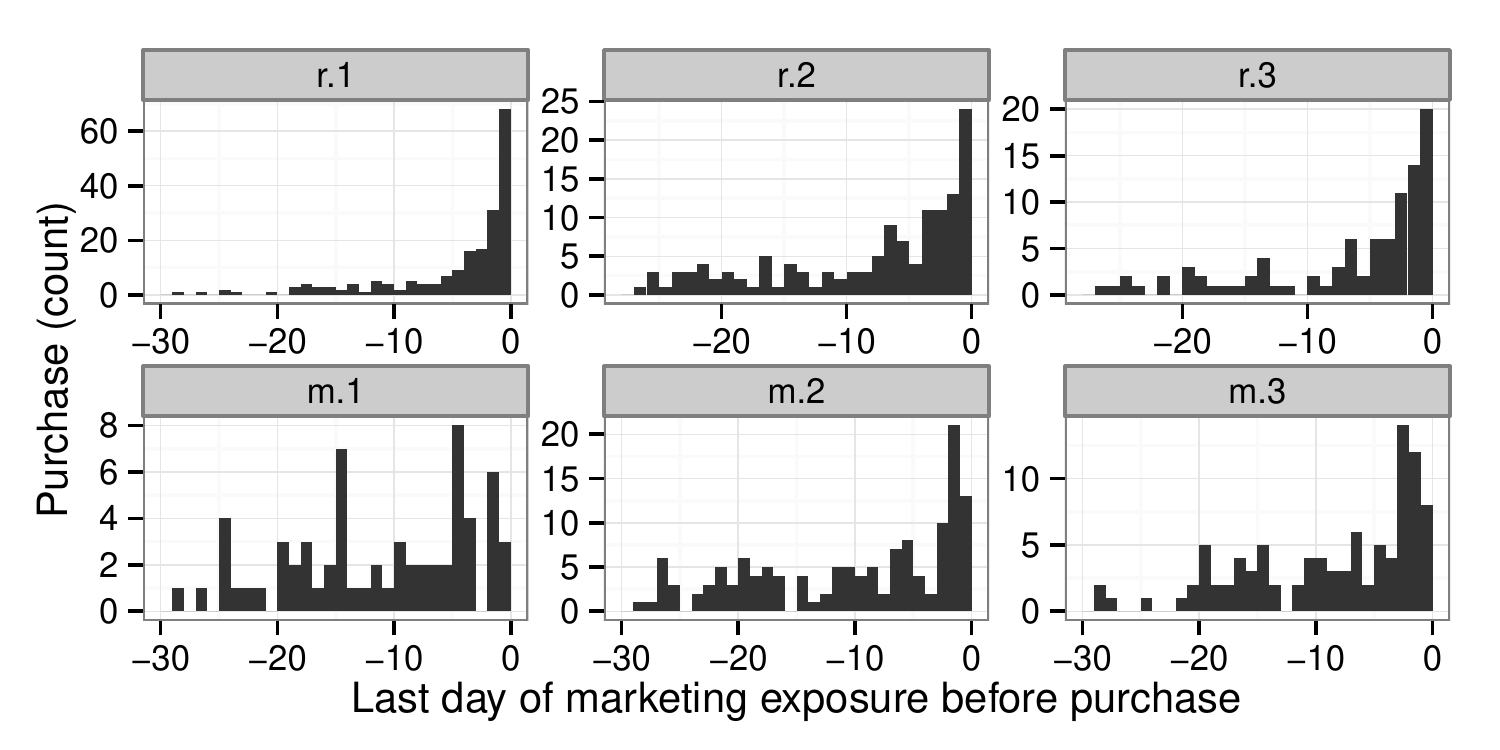}
	\caption{Product B}
\end{subfigure}
\caption{Evidence of time-lagged marketing effects. Each cell represents a different marketing touch. }\label{fig:evidence}
\end{figure}

One of the main motivations for designing DPM was the time dependency of marketing effects. 
The positive effects of past marketing touches can be easily illustrated.
Figure~\ref{fig:evidence} shows the histogram of the day of the last marketing exposure before purchase.
The x-axis represents time at a day resolution, where $x=0$ is the time of purchase.
As can be seen, past marketing events are correlated to purchases.
Noticeably, the temporal effects show exponential decay, which suggest an underlying autoregressive process.

\subsection{Baseline Models}

We build three logistic regression models using lagged variables as follows:
\begin{small}
\begin{align*}
\text{(glm)}:\quad\quad& E[y_{t+1} \mid \mathbf{r}_{t}, \mathbf{m}_{t}] = \text{logit}^{-1}(c +  \boldsymbol{\alpha}_0^\top \mathbf{r}_t + \boldsymbol{\beta}_0^\top \mathbf{m}_t)\\
\text{(glm.lag1)}:\quad\quad& E[y_{t+1} \mid \mathbf{r}_{t:(t-1)}, \mathbf{m}_{t:(t-1)}] = \text{logit}^{-1}(c +  \sum_{l = 0}^{1} \boldsymbol{\alpha}_l^\top \mathbf{r}_{t-l} + \sum_{l = 0}^{1} \boldsymbol{\beta}_l^\top \mathbf{m}_{t-l} )
\\
\text{(glm.lag2)}:\quad\quad& E[y_{t+1}  \mid \mathbf{r}_{t:(t-2)}, \mathbf{m}_{t:(t-2)}] = \text{logit}^{-1}(c + \sum_{l = 0}^{2} \boldsymbol{\alpha}_l^\top \mathbf{r}_{t-l} +  \sum_{l = 0}^{2}\boldsymbol{\beta}_l^\top \mathbf{m}_{t-l} )
\end{align*}
\end{small}
where $\boldsymbol{\alpha}_l$ and $\boldsymbol{\beta}_l$ represent the effects for the $l$-lagged marketing touches.
These models are estimated using the \texttt{glm} method in \texttt{R} 2.15.3.
Since our datasets have highly imbalanced target ratios, the parameters of these models with lagged variables, especially \texttt{glm.lag1}, sometimes do not converge. 
We do not show the models with deep lagged variables $(l>2)$, since those models are much harder to estimate.

\subsection{Estimated Parameters}

\begin{figure}
\center
\begin{subfigure}[b]{1\textwidth}
	\includegraphics[width=1\textwidth]{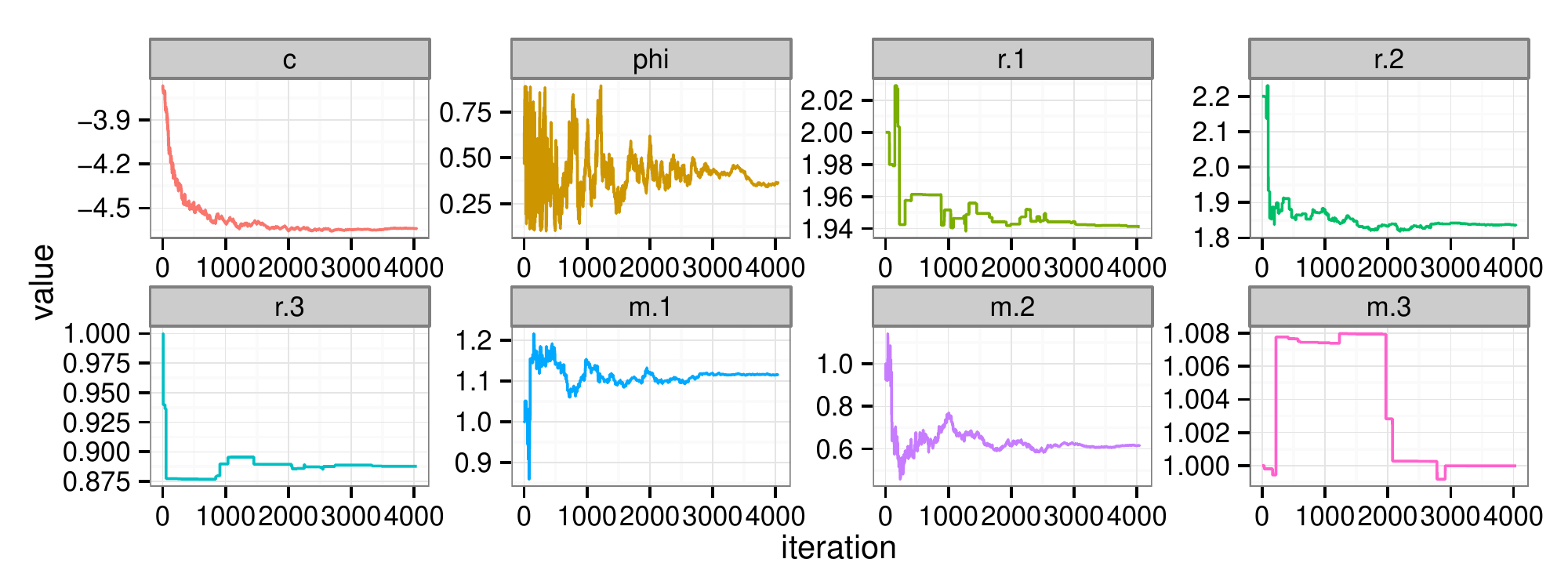}
	\caption{Product A}
\end{subfigure}

\begin{subfigure}[b]{1\textwidth}
	\includegraphics[width=1\textwidth]{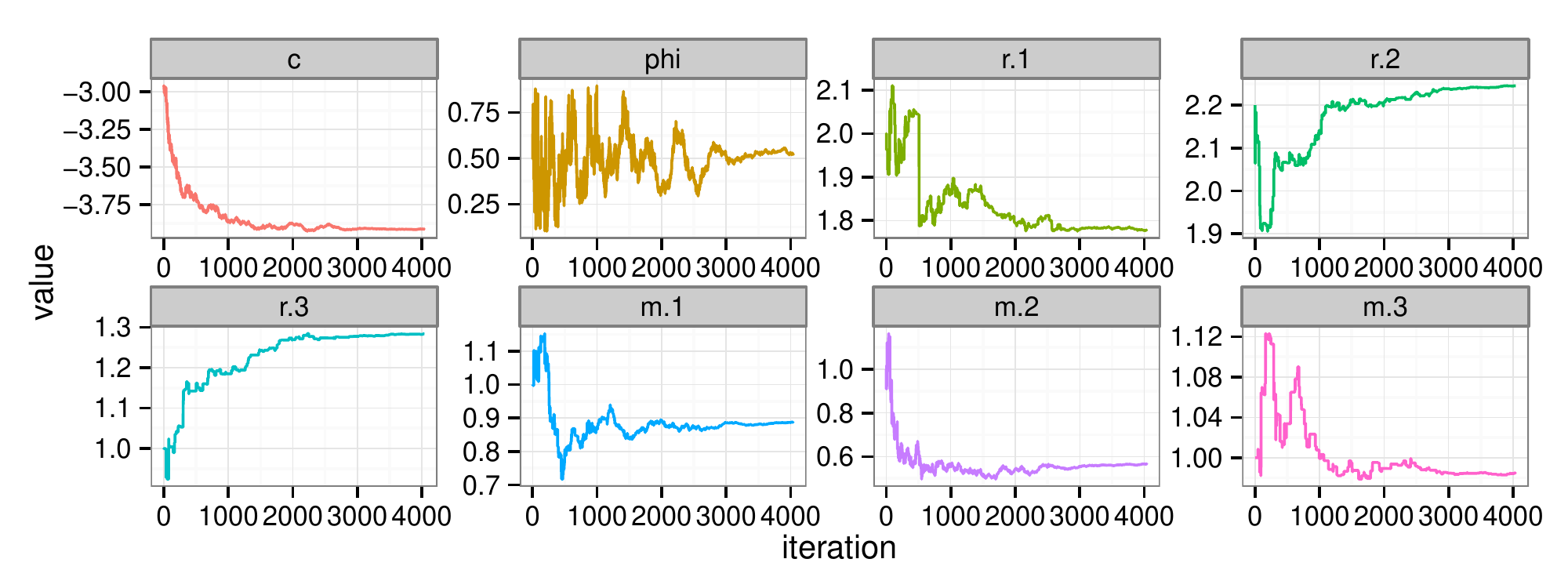}
	\caption{Product B}
\end{subfigure}
\caption{SGD estimation of the DPM parameters.}\label{fig:sgdpm}
\end{figure}

We estimate the parameters of DPM using the SGDPM algorithm.
Figure~\ref{fig:sgdpm} shows the estimated parameters over iterations.
Recall that each iteration in SGDPM represents a data block of one customer.
The customers who purchased the products are selected with a higher probability, since otherwise SGDPM may not see any positive examples before it converges.
As can be seen, the parameters almost converge after visiting approximately two thousand customers.
The damping parameters for Product A and Product B are 0.36 and 0.53, respectively. 
Marketing effects diminish by about half in the next day, but they still positively affect purchases.

\begin{figure}
\center
\includegraphics[width=0.9\textwidth]{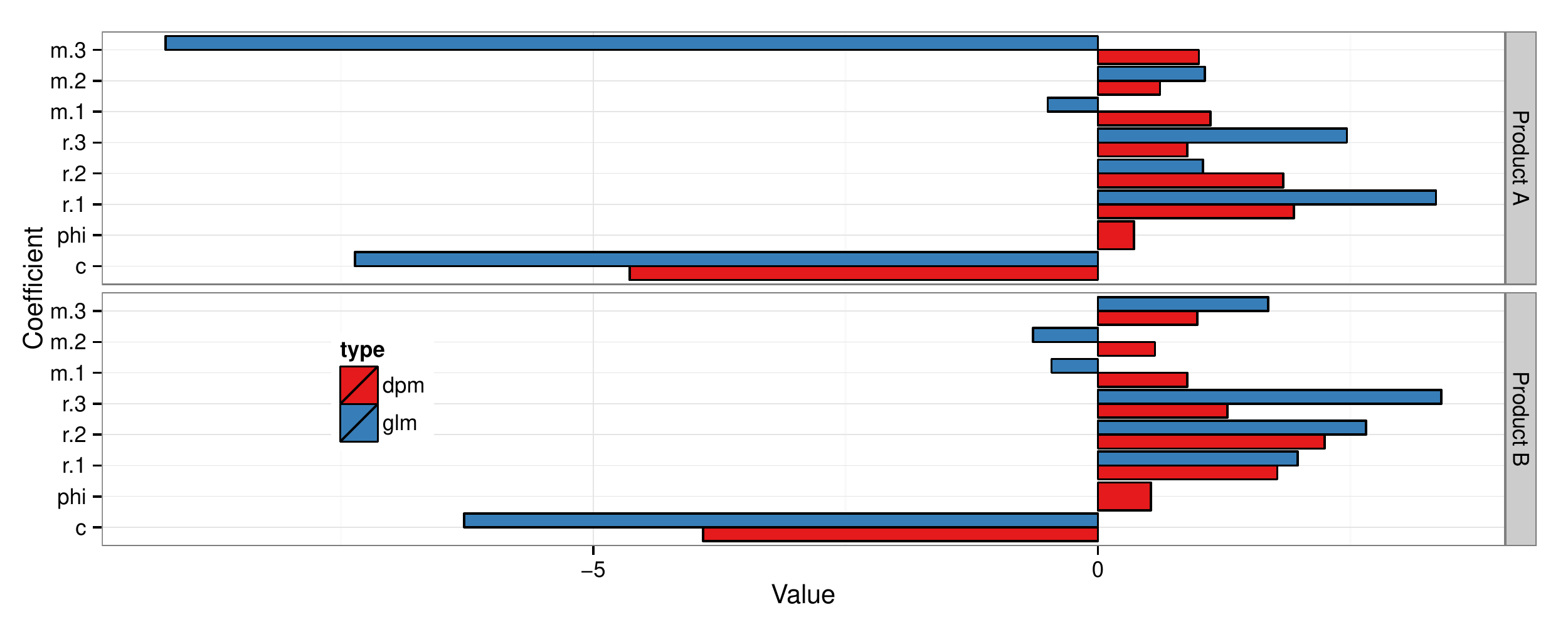}
\caption{Estimated Parameters from DPM and GLM.}\label{fig:coeff}
\end{figure}

The estimated parameters from the baseline models are significantly different from the DPM parameters.
Figure~\ref{fig:coeff} compares the coefficients from two different models: DPM and \texttt{glm}.
As can be seen in the figure, \texttt{glm} outputs negative weights for some of the marketing effects e.g. $\beta_{01} < 0 $ and $\beta_{03} < 0 $ for Product A, and $\beta_{01} < 0 $ and $\beta_{02} < 0 $ for Product B.
These negative coefficients mainly appear on the targetable marketing touches $\mathbf{m}_t$.
A viable explanation for this is that the company is already targeting a specific group of customers whose purchase rates are lower than normal population.
On the other hand, DPM shows all positive coefficients for the marketing touches.
DPM tracks customers' marketing histories, and adjusts the current offsets $x_t$ for purchases.
Thus, the coefficients from DPM are more robust for sampling biases than the baseline models.

Table~\ref{tab:productb_coeff} shows the estimated parameters of the three baseline models for Product B.
The negative effects have higher p-values compared to the positive effects.
For example, the p-value of \texttt{glm.lag1} $\beta_{12}$ is 0.974. 
Some of the lagged variables have significant effects on purchases, which supports our claim that past marketing touches affect purchase behaviors. 
These GLM models put more weights on the semi-targetable marketing touches than the targetable marketing touches.
Unless the effects of the targetable marketing touches are estimated using control and test groups, these results do not provide insights for building actionable marketing strategies.

\begin{table}\caption{Estimated Parameters for Product B from Generalized Linear Models. }\label{tab:productb_coeff}
\begin{small}
\begin{center}
\begin{tabular}{ c  c  c  c  c  c  c   }
\hline
& \multicolumn{2}{c  }{\textbf{glm}} & \multicolumn{2}{c  }{\textbf{glm.lag1}} & \multicolumn{2}{c  }{\textbf{glm.lag2}} \\ \hline
Param. & Estimate  & p-value & Estimate  & p-value & Estimate  & p-value \\\hline\hline
$\alpha_{01}$ & 2.02031 & $<$ 2e-16 & 1.90741 & $<$ 2e-16 & 1.84673 &  $<$ 2e-16\\
$\alpha_{02}$ & 2.74625 &  $<$ 2e-16  & 2.66615 &$<$ 2e-16 & 2.59654 & $<$ 2e-16  \\
$\alpha_{03}$ & 3.16096 & $<$ 2e-16   & 3.14794  & $<$ 2e-16& 3.18083 & $<$ 2e-16 \\\hline
$\beta_{01}$ & -0.59591 & 0.185  & -0.59042 & 0.18922& -0.59176 & 0.188291 \\
$\beta_{02}$ & -0.32632 & 0.308 & -0.31584  & 0.32377&-0.33462 &  0.296802\\
$\beta_{03}$ &  1.30361 & 1.74e-05 & 1.25721 & 4.05e-05& 1.25432 & 4.52e-05 \\\hline\hline
$\alpha_{11}$ & - & - & 0.66684  &0.00018 &0.51507 & 0.007232 \\
$\alpha_{12}$ & - & - & 0.60640&0.05751 & 0.55132 &  0.086354\\
$\alpha_{13}$ & - & - & 1.63553  & 5.69e-05& 1.65591 &  4.32e-05\\\hline
$\beta_{11}$ & - & -  & -1.25127& 0.21076& -1.24799 & 0.212039 \\
$\beta_{12}$ & -  & -  & -0.01249&  0.97401 & -0.04239 &  0.911988\\
$\beta_{13}$ & - & - & 1.21296 & 0.01544 & 1.10564 &  0.029452\\\hline\hline
$\alpha_{21}$ & -  &- & -& -& 0.52325 &  0.013828\\
$\alpha_{22}$ &-  &- & -& -& 0.52655 & 0.149112 \\
$\alpha_{23}$ &- & -& -& -&1.53267 &  0.000412\\\hline
$\beta_{21}$ &- & -& -& -& -0.11958 & 0.836673 \\
$\beta_{22}$ &- &- &- &- & 0.62343&  0.025465\\
$\beta_{23}$ &- &- &- &- & 1.71137& 2.53e-05 \\\hline
\end{tabular}
\end{center}
\end{small}
\end{table}

\subsection{Predictive Performance}

\begin{figure}
\center
\begin{subfigure}[b]{0.45\textwidth}
	\includegraphics[width=1\textwidth]{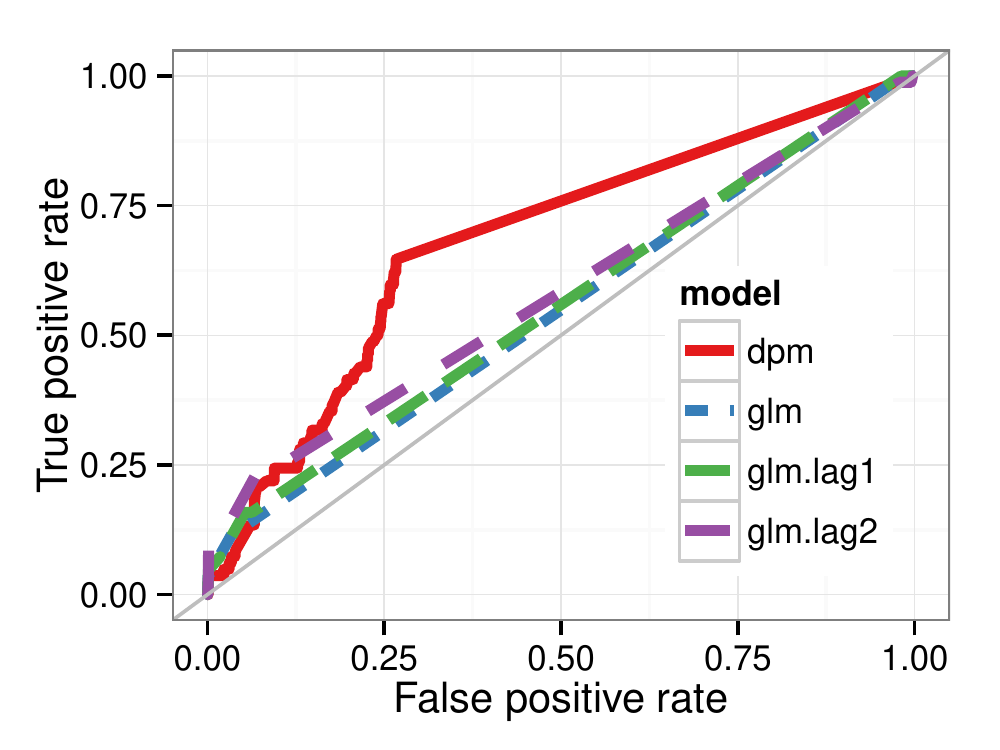}
	\caption{Product A}
\end{subfigure}
\begin{subfigure}[b]{0.45\textwidth}
	\includegraphics[width=1\textwidth]{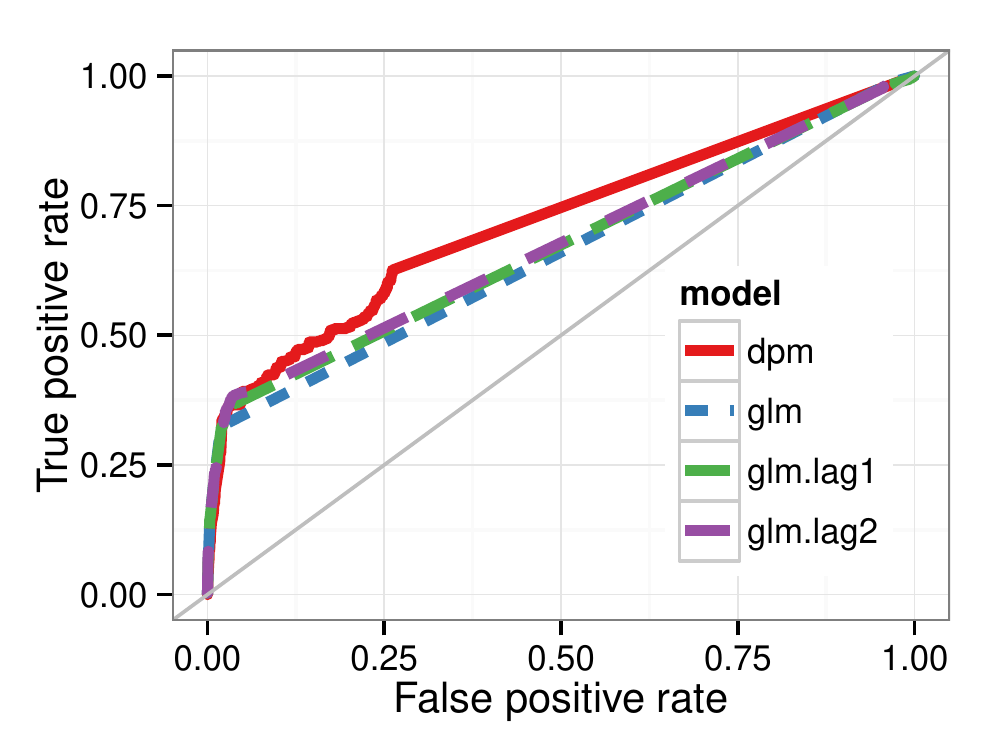}
	\caption{Product B}
\end{subfigure}
\caption{Receiver Operating Characteristic curves from the test datasets.}\label{fig:roc}
\end{figure}

We measure the predictive performances of DPM and the baseline models using hold-out test datasets.
The customers in the test datasets do not overlap with the customers in the training datasets.
The performances were measured using Receiver Operating Characteristics (ROC) curves.
Figure~\ref{fig:roc} shows the measured ROC curves for both Product A and B.
As can be seen, the DPM's areas under the curves are significantly higher than those of the baseline models.

Although their performances are comparable when False Positive Ratios are close to zero (FPR $\approx$ 0), the baseline models cannot capture the purchase  probabilities of the customers who had (latent) intentions of purchasing the products.
In other words, the logistic regression models are trained directly on the observed variables, and there is no latent variable involved.
On the other hand, DPM, a latent-variable time series model, simultaneously estimates both the intention of purchasing the products and the parameters of the model. 
The latent intentions of purchasing are captured through marketing touches and their corresponding responses: the customers who clicked display ads, who received referrals, or who positively responded to promotion phone calls.
By modelling these latent intentions, DPM achieves better performance curves than traditional lagged variable model.

\begin{figure}[h]
\center
\includegraphics[width=0.9\textwidth]{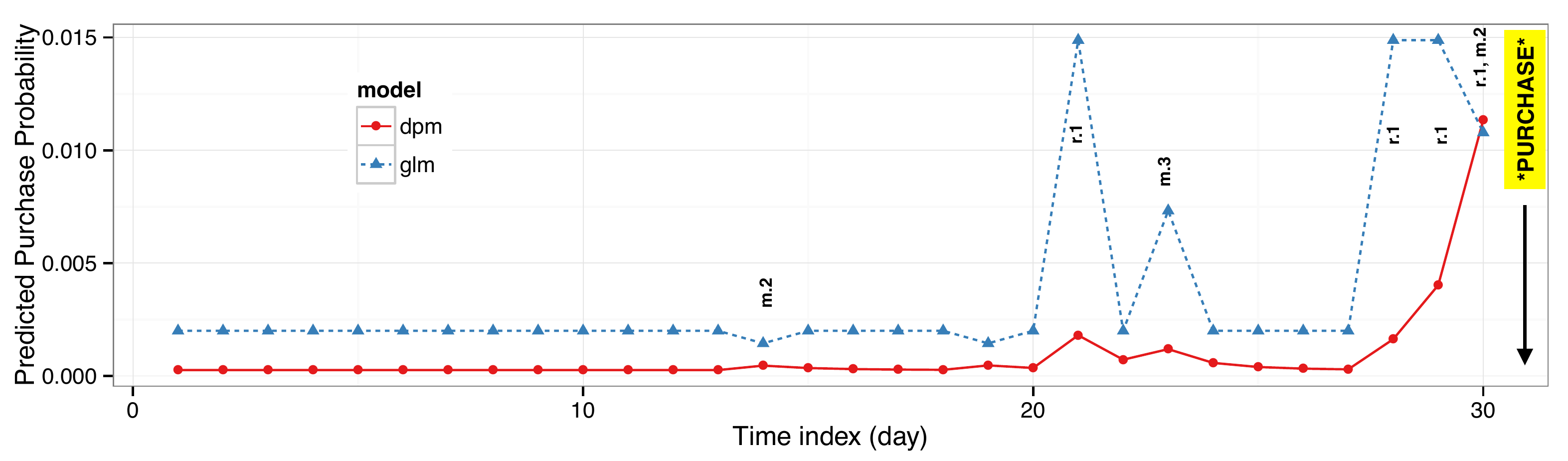}
\caption{Visualization of dynamic purchase probability.}\label{fig:illust}
\end{figure}

The visualization of the latent variable provides a different perspective on purchase behaviors.
Figure~\ref{fig:illust} shows the dynamic purchase probabilities from two models of a customer.
A customer's propensity to purchase propagates and accumulates over time.
The predicted purchase probability of DPM shows the process of building up a purchase decision.
On the 28th and 29th day, the customer received and responded to the $r_1$ marketing touch, and on the 30th day, he was targeted by the $m_2$ marketing touch.
Through the process of marketing touches and responses, his propensity to purchase Product B has gradually increased.
On the other hand, the baseline model (\texttt{glm}) does not capture this cumulative process, and it actually provides a lower probability score on the purchase day than the day before.

\section{Concluding Remarks}\label{sec:conclusion}

In this paper, we proposed a state space model, DPM, to answer three main challenges in direct marketing: 
\textit{which channel} to use, \textit{which offer} to make, and \textit{when} to offer.
DPM is a latent variable time series model that utilizes demographics, and both marketing and purchase histories of a customer. 
To estimate the parameters of the model,  a new statistical methodology, SGDPM, was developed.
This methodology combines state-space models with a stochastic gradient descent approach, resulting in fast estimation of the model coefficients from big data.
The experimental results using a real dataset showed that DPM can effectively forecast the time of purchase.

We used only six marketing variables to focus the modeling idea of DPM: three semi-targetable and three targetable marketing touches.
In practice, combining demographic information with DPM can significantly improve the predictive performance \citep{Risselada2013}.
For example, marketing demographic segments can be used in a multi-level modeling approach; each segment would have slightly different parameters.
Social network information \citep{Palmer2009} can also leverage the predictive performance of the model.
Specifically, information about customers' important life events, such as a marriage or a new baby, can be easily used to extend the model as follows:
\begin{align*}
s_{t+1}^i = c + \phi x_{t}^i + \boldsymbol{\alpha}^\top \mathbf{r}_t^i + \boldsymbol{\beta}^\top \mathbf{m}_{t}^i  + \boldsymbol{\psi}^\top \underbrace{\mathbf{e}_t^i }_{\mathclap{ \text{life events}}}
\end{align*}
where $\mathbf{e}_t^i$ represents a vector of life events. 
Some life events, such as having a family, may increase the propensity of buying financial products ($\psi > 0$), whereas a better deal from an rival company would decrease the propensity ($\psi < 0$).

Although we showed the predictive performance of DPM using hold-out datasets,  the true predictive performance of the model should also be confirmed by thorough A/B-testing or randomized controlled experiments.
The scores from DPM and other existing models can also be combined to increase purchase rates.
To build a personalized marketing strategy, customer feedback needs to be fully utilized.
An IT infrastructure for daily monitoring of customers should be the first step for these types of dynamic models in practice.
Extensions of our approaches and further discussions on implementation methods are left as future work.

\bibliographystyle{abbrvnat}
\bibliography{dpm} 

\end{document}